\documentstyle[preprint,aps]{revtex}
\tighten
\begin{document}
\preprint{UCONN-95-05}
\draft
\title{Implementing Gauss's law in Yang-Mills theory and QCD}
\author{Mario Belloni$^{\dagger}$, Lusheng Chen$^{\ddagger}$, and
Kurt Haller$^{\ast}$}
\address{Department of Physics, University of Connecticut,
Storrs, Connecticut 06269}
\date{October 6, 1995}
\maketitle
\begin{abstract}
We construct a transformation that transforms perturbative states
into states that implement Gauss's law for `pure gluonic' Yang-Mills
theory and QCD.  The fact that this transformation is not and cannot
be unitary has special significance.  Previous work has shown that
only states that are unitarily equivalent to perturbative states
necessarily give the same S-matrix elements as are obtained with
Feynman rules.
\end{abstract}
\vskip 2.3cm

\vskip 7.5cm

\line(1,0){430}

$^{\dagger}$ e-mail address: mario@main.phys.uconn.edu

$^{\ddagger}$ e-mail address: chen@main.phys.uconn.edu

$^{\ast}$ e-mail address: KHALLER@UCONNVM.UCONN.EDU

In earlier work, one of us (KH) quantized Yang-Mills theory in the
temporal ($A^a_0=~\!\!0$) gauge and formulated the constraint that
implements Gauss's law by selecting an appropriate subspace for the
dynamical time evolution of state vectors\cite{khymtemp}.
One objective of Ref.~\cite{khymtemp} was to compare the
implications of Gauss's law when it is imposed on the
non-Abelian Yang-Mills (YM) theory and Quantum Chromodynamics
(QCD) with its role in QED. Despite great similarities between the
Abelian and the non-Abelian theories, the inclusion of gauge fields
in the non-Abelian charge density is responsible for important
differences between QED and QCD~\cite{khymtemp,khqedtemp}.
Not only is it far more difficult to construct states that
implement Gauss's law in non-Abelian gauge theories than in QED;
it is also much more important to use states that implement
Gauss's law in evaluating S-matrix elements in YM theory
and QCD than it is in QED. \bigskip

In gauge theories, it is standard practice to use Feynman rules
in perturbative calculations; these rules implicitly use charged
particle states that do {\em not} obey Gauss's law. In the
evaluation of S-matrix elements in QED, perturbative states
that do not implement Gauss's law may be safely substituted
for states that do implement it. This has been shown to be due
to the fact that unitary transformations suffice to construct
the latter states from the former~\cite{khqedtemp,khelqed}.
But this unitary equivalence does not extend to the non-Abelian
gauge theories. The validity of perturbative calculations based
on Dirac spinor quarks and free gluons may therefore require
qualifications in YM theory and QCD that are not needed in QED.
Furthermore, the question has been raised whether
the proper implementation of Gauss's law in non-Abelian gauge
theories might have significant implications for the confinement
of colored states and the conjectured requirement that only
color singlets can be asymptotic scattering states\cite{khymtemp}.
\bigskip

In this paper, we construct states that obey the non-Abelian
Gauss's law in `pure gluonic' YM theory and QCD.  Our program is
based on the construction of a transformation ${\cal{T}}$
(which must be non-unitary) that transforms perturbative states
$|a\rangle$ --- in the first instance
the perturbative vacuum state $|0\rangle$ --- into states that
satisfy Gauss's law, and that continue to satisfy Gauss's law even
after dynamical time evolution. Unlike James and Landshoff,
who had to require matrix elements of a
non-terminating progression of powers of the ``Gauss's
law operator'', $J^a_0-\partial_{i}E^a_{i},$ to vanish
in order to obtain states that implement Gauss's law~\cite{lands},
we find that we do not need to apply progressively
escalating powers of projection operators to achieve our objective.
\bigskip

As in Refs.~\cite{khymtemp,khqedtemp}, we represent the
transverse gauge fields and their adjoint momenta as
\begin{equation}
A_{Ti}^{a}({\bf{r}}) = \sum_{{\bf{k}}; s=1,2}
\frac{\epsilon_{i}^{s}({\bf{k}})}{\sqrt{2k}}\,
[a_{s}^{a}({\bf{k}})e^{+i{\bf{k}}\cdot{\bf{r}}} +
a_{s}^{a\dagger}({\bf{k}})e^{-i{\bf{k}}\cdot{\bf{r}}}]\;,
\label{eq:atrans}
\end{equation}
and
\begin{equation}
\Pi_{Ti}^{a}({\bf{r}}) = \sum_{{\bf{k}}; s=1,2}
-i\epsilon_{i}^{s}({\bf{k}})\sqrt{k/2}\,[a_{s}^{a}({\bf{k}})
e^{+i{\bf{k}}\cdot{\bf{r}}} - a_{s}^{a\dagger}({\bf{k}})
e^{-i{\bf{k}}\cdot{\bf{r}}}]\;,
\end{equation}
where $a_{s}^{a\dagger}({\bf{k}})$ and $a_{s}^{a}({\bf{k}})$
represent `standard' creation and annihilation operators, respectively,
for gluons (or --- with the Lie group indices removed --- photons) of
helicity $s.$  The
longitudinal fields are represented in terms of ghost excitation
operators, in the form
\begin{equation}
A_{Li}^{a}({\bf{r}}) =
\sum_{{\bf{k}}}\frac{k_{i}}{2k^\frac{3}{2}}\,[a_{R}^{a}({\bf{k}})
e^{+i{\bf{k}}\cdot{\bf{r}}}
+a_{R}^{a\star}({\bf{k}})e^{-i{\bf{k}}\cdot{\bf{r}}}]\;,
\label{eq:along}
\end{equation}
and
\begin{equation}
\Pi_{Li}^{a}({\bf{r}}) = \sum_{{\bf{k}}}
\frac{-ik_{i}}{\sqrt{k}}\,[a_{Q}^{a}({\bf{k}})
e^{+i{\bf{k}}\cdot{\bf{r}}} -
a_{Q}^{a\star}({\bf{k}})e^{-i{\bf{k}}\cdot{\bf{r}}}]\;.
\label{eq:pilong}
\end{equation}
The ghost excitation operators obey the commutation rules
$[a_{Q}^{a}({\bf{k}}),a_{R}^{b\star}({\bf{k^\prime}})] =
\delta_{ab}\delta_{{\bf{k,k^\prime}}},$
and $[a_{R}^{a}({\bf{k}}),a_{Q}^{b\star}({\bf{k^\prime}})] =
\delta_{ab}\delta_{{\bf{k,k^\prime}}},$
with all other commutators vanishing. We define the
``Gauss's law operator'' ${\cal{G}}^{a}({\bf{r}})$
\begin{equation}
{\cal{G}}^{a}({\bf{r}})=\partial_{i}{\Pi}^a_{i}({\bf{r}})
+J^a_0({\bf{r}})=-\partial_{i}E^a_{i}({\bf{r}})+J^a_0({\bf{r}})\,;
\end{equation}
where $J_{0}^{a}({\bf{r}}) = g\,f^{abc}A_{i}^{b}({\bf{r}})\,
\Pi_{i}^{c}({\bf{r}}).$
${\cal{G}}^{a}({\bf{r}})$ can also conveniently be represented in
the form
\begin{equation}
{\cal{G}}^{a}({\bf{r}}) = \frac{1}{2}\sum_{{\bf{k}}}\,
[\Omega^{a}({\bf{k}}) +
\Omega^{a\star}(-{\bf{k}})]\,e^{+i{\bf{k}}\cdot{\bf{r}}}\;,
\label{eq:gomega}
\end{equation}
where\footnote{$\Omega^{a}({\bf{k}})$ ---
and $\Omega^{a\star}({\bf{k}})$ --- in this work and in Ref.~\cite{khymtemp}
differ by a normalization factor of  $2k^{3/2}$.}
\begin{equation}
\Omega^{a}({\bf{k}}) = 2k^{\frac{3}{2}}a_{Q}^{a}({\bf{k}}) +
J_{0}^{a}(\bf{k})\;.
\label{eq:omega}
\end{equation}
In earlier work\cite{khymtemp,khqedtemp}, it was demonstrated that
Gauss's law and the gauge choice, $A^a_0=0,$ could be implemented
by imposing
\begin{equation}
\Omega^{a}({\bf{k}}) |\nu\rangle=0\;
\label{eq:subcon}
\end{equation}
on a set of states, $\{|\nu\rangle\}.$   $\Omega^{a}({\bf{k}})$ --- and its
adjoint
$\Omega^{a\star}({\bf{k}})$ --- commute with the Hamiltonians for YM
theory and QCD, as well as (with the Lie group index $a$ removed)
for QED. Gauss's law, once imposed by
this method, therefore is unaffected by time evolution and remains
permanently intact. \bigskip

In QED, the Noether current commutes with the gauge field,
because the photons couple to, but do not carry the electric charge.
$\Omega({\bf{k}})$ therefore commutes with $\Omega^{\star}({\bf k}),$
mirroring the commutation rule between $a_{Q}({\bf k})$ and
$a_{Q}^{\star}({\bf k}).$
That makes it possible to establish a unitary equivalence
between the set of states $\{|\nu\rangle\}$
--- the solutions of Eq.~(\ref{eq:subcon}) --- and the set of
states that solve $a_{Q}({\bf{k}}) |n\rangle=0.$ We are able to
exploit this unitary equivalence to explicitly construct the
states in $\{|\nu\rangle\},$ and to reformulate QED as a theory
of charged particles that obey Gauss's law --- that therefore
carry their Coulomb field with them --- and that interact with each
other and with transversely polarized propagating photons
\cite{khqedtemp,khelqed}.  In YM theory and QCD, however, the
commutation rules among the components of $\Omega^{a}({\bf{k}})$
are quite different from the commutation rules among the corresponding
$a_{Q}^{a}({\bf{k}}).$  The latter `ghost' annihilation operators
commute with each other and with
all their conjugate $a_{Q}^{a\star}({\bf{k}}).$  However, the commutation
rules for  $\Omega^{a}({\bf{k}})$ and $\Omega^{a\star}({\bf{k}})$
follow the $SU(N)$ Lie algebra\cite{khymtemp}.
It is therefore impossible to construct a unitary transformation
that transforms $\Omega^{a}({\bf{k}})$ into $a_{Q}^{a}({\bf{k}}).$
This difference between QED on the one hand and YM theory and QCD
on the other, precludes the use of unitary transformations to
construct the set of physical states $\{|\nu\rangle\}$ from the
perturbative states $\{|n\rangle\}$ in the non-Abelian
gauge theories; and that fact accounts for major differences
between QED and the non-Abelian YM theory and QCD. \bigskip

In formulating a procedure for constructing the state
$|{\nu}_0\rangle$ in non-Abelian gauge theories, we note that
whereas $\Omega^{a}({\bf{k}}) |\nu\rangle=0$ and
$\Omega^{a\star}({\bf{k}}) |\nu\rangle=0$
are two wholly independent conditions for QED --- and the former alone
suffices to define the Fock space for this Abelian theory --- these two
conditions are not independent for YM theory and
QCD. In YM theory and QCD, Eq.~(\ref{eq:subcon}) requires that
$\Omega^{a\star}({\bf{k}}) |\nu\rangle=0$
too\cite{khymtemp,goldjack,jackiw}.  The appropriate
condition for imposing Gauss's law in YM theory therefore is not
Eq.~(\ref{eq:subcon}), but
\begin{equation}
[\Omega^{a}({\bf{k}})+\Omega^{b\star}({\bf{-k}})]|{\nu}\rangle=0\;.
\end{equation}
We will express this condition as
\begin{equation}
[b_{Q}^{a}({\bf{k}}) + J_{0}^{a}({\bf{k}})]|\nu\rangle = 0\;,
\label{eq:newsubcon}
\end{equation}
where we define
\begin{equation}
 b_{Q}^{a}({\bf{k}}) =
k^{\frac{3}{2}}[a_{Q}^{a}({\bf{k}}) +
a_{Q}^{a\star}(-{\bf{k}})]\;.
\label{eq:bq}
\end{equation}
\bigskip

We will transform the perturbative vacuum state
$|0\rangle$ --- the state that is annihilated by $a_{s}^{a}({\bf{k}}),$
$a_{Q}^{a}({\bf{k}})$ and $a_{R}^{a}({\bf{k}})$ --- into a state
$|{\nu}_0\rangle$ that implements Gauss's law.  We represent
$|{\nu}_0\rangle$ as a product of two operators acting
on the perturbative vacuum $|0\rangle,$ in the form
\begin{equation}
|{\nu}_0\rangle={\Psi}\;{\Xi}\;|0\rangle\;,
\label{eq:Nustate}
\end{equation}
where the operator product ${\Psi}\,{\Xi}\,$ represents a non-unitary
transformation operator. We also define a state
$|{\phi_0}\rangle={\Xi}\;|0\rangle,$ so that
\begin{equation}
b_Q^a({\bf k})|{\phi_0}\rangle=0\;.
\label{eq:phisub}
\end{equation}
Eq.~(\ref{eq:phisub}) is satisfied by ${\Xi}=
\exp\{-\sum_{\mbox{\boldmath $\kappa$}}
a_{R}^{c\,\star}(\mbox{\boldmath $\kappa$})
a_{Q}^{c\,\star}(-\mbox{\boldmath $\kappa$})\}$,
 as is confirmed by the observation that
$[a_Q^a({\bf k}),\,{\Xi}\,]=-\,a_{Q}^{a\star}({\bf -k})\,{\Xi}$
and $[a_Q^{a\star}({\bf k}),\,{\Xi}\,]=0.$
The resulting state $|{\phi_0}\rangle$ is not normalizable --- it is
essentially the ``Fermi'' vacuum state~\cite{fermi}, which is not
commonly used in QED, but reappears here in the non-Abelian theory. \bigskip

The construction of ${\Psi}$ involves solving the equation
$\{b_{Q}^{a}({\bf{k}}) + J_{0}^{a}({\bf{k}})\}\,{\Psi}\,|{\phi_0}\rangle=
{\Psi}b_{Q}^{a}({\bf{k}})|{\phi_0}\rangle$, or
equivalently,
\begin{equation}
[b_{Q}^{a}({\bf{k}}),\,{\Psi}\,]=-J_{0}^{a}({\bf{k}})\,{\Psi}+B_Q^{a}\;,
\label{eq:psicom}
\end{equation}
where $B_Q^{a}$ represents any operator product with $b_Q^{a}$ on its
extreme right-hand side,
so that $B_Q^{a}|{\phi_0}\rangle =0.$ Eq.~(\ref{eq:psicom})
is an operator differential equation, in which the commutator
plays the role of a generalized derivative.  We introduce the following
notation for the constituent parts from which ${\Psi}$ will be assembled:
\begin{equation}
a_i^\alpha({\bf{r}}) = A_{Ti}^{\alpha}({\bf{r}})\;,
\label{eq:bookai}
\end{equation}
\begin{equation}
x_i^\alpha({\bf{r}}) = A_{Li}^{\alpha}({\bf{r}})\;,
\label{eq:bookxi}
\end{equation}
and
\begin{equation}
{\cal{X}}^\alpha({\bf{r}}) =
[{\textstyle\frac{\partial_i}{\partial^2}}A_i^\alpha({\bf{r}})]\;,
\end{equation}
where, $[a_i^\alpha({\bf{r}})+x_i^\alpha({\bf{r}})]=
A_{i}^{\alpha}({\bf{r}}),$
and since $A_{Li}^{\alpha}({\bf{r}}) =
\partial_i[{\textstyle\frac{\partial_j}{\partial^2}} A_j^\alpha({\bf{r}})]\,
,\;x_i^\alpha({\bf{r}})=\partial_i{\cal{X}}^\alpha({\bf{r}})$.
We also need to define the combination
\begin{equation}
{\cal{Q}}_{(\eta)i}^{\beta}({\bf{r}}) =
[a_i^\beta ({\bf{r}})+
{\textstyle\frac{\eta}{\eta+1}}x_i^\beta({\bf{r}})]\;,
\label{eq:bookaiQ}
\end{equation}
with $\eta$ integer-valued.
\bigskip

We will also use the preceding operators to form the following
composite operators:
\begin{equation}
\psi^{\gamma}_{(1)i}({\bf{r}})= \,f^{\alpha\beta\gamma}\,
{\cal{X}}^\alpha({\bf{r}})\;{\cal{Q}}_{(1)i}^{\beta}({\bf{r}})
=\,f^{\alpha\beta\gamma}\,
{\cal{X}}^\alpha({\bf{r}})\,[ a_i^\beta({\bf{r}}) +
{\textstyle\frac{1}{2}}x_i^\beta({\bf{r}}) ]\;,
\label{eq:psi1def}
\end{equation}
\begin{equation}
\psi^{\gamma}_{(2)i}({\bf{r}})=-f^{\alpha\beta b}\,f^{b\delta\gamma}\,
{\cal{X}}^\alpha({\bf{r}})\,{\cal{Q}}_{(2)i}^{\beta}({\bf{r}})\,
{\cal{X}}^\delta({\bf{r}})
=-f^{\alpha\beta b}\,f^{b\delta\gamma}\,
{\cal{X}}^\alpha({\bf{r}})\,[a_i^\beta ({\bf{r}})+
{\textstyle\frac{2}{3}}x_i^\beta({\bf{r}})]\,{\cal{X}}^\delta({\bf{r}})\;,
\label{eq:psi2def}
\end{equation}
and the general $\eta$-th order term
\begin{equation}
\psi^{\gamma}_{(\eta)i}({\bf{r}})= \,(-1)^{\eta-1}\,
f^{\vec{\alpha}\beta\gamma}_{(\eta)}\,
{\cal{R}}^{\vec{\alpha}}_{(\eta)}({\bf{r}})\;
{\cal{Q}}_{(\eta)i}^{\beta}({\bf{r}})\;,
\label{eq:psindef2}
\end{equation}
in which
\begin{equation}
{\cal{R}}^{\vec{\alpha}}_{(\eta)}({\bf{r}})=
\prod_{m=1}^\eta{\cal{X}}^{\alpha[m]}({\bf{r}})\;,
\label{eq:XproductN}
\end{equation}
and
\begin{equation}
f^{\vec{\alpha}\beta\gamma}_{(\eta)}=f^{\alpha[1]\beta b[1]}\,
\,f^{b[1]\alpha[2]b[2]}\,f^{b[2]\alpha[3]b[3]}\,\cdots\,
\,f^{b[\eta-2]\alpha[\eta-1]b[\eta-1]}f^{b[\eta-1]\alpha[\eta]\gamma}\;,
\label{eq:fproductN}
\end{equation}
where $f^{\vec{\alpha}\beta\gamma}_{(1)}\equiv f^{\alpha\beta\gamma}$.\bigskip

We also define the composite operator
\begin{equation}
{\cal{A}}_1=ig{\int}d{\bf r}\;\psi^{\gamma}_{(1)i}({\bf{r}})\;
\Pi_{i}^{\gamma}({\bf{r}})\;,
\label{eq:defscra1}
\end{equation}
which is useful because it has the important property that its commutator
with $b_Q^a({\bf{k}}),$
\begin{eqnarray}
[b_Q^a({\bf{k}}),\,{\cal{A}}_1]\;&&=
-g\,f^{a\beta\gamma}{\int}d{\bf{r}}\;e^{-i{\bf{k\cdot r}}}\;
 [ a_i^\beta({\bf{r}}) + x_i^\beta({\bf{r}}) ]\;\Pi_i^\gamma({\bf{r}})
\nonumber\\
&&-{\textstyle\frac{g}{2}}\,f^{a\alpha\gamma}\;{\int}d{\bf{r}}\;
e^{-i{\bf{k\cdot r}}}\;
{\cal{X}}^\alpha\;[\partial_i\Pi_i^\gamma({\bf{r}})]\;,
\label{eq:thetabqcom}
\end{eqnarray}
generates $-J_{0}^{a}({\bf{k}})$  when it acts on the ``Fermi"
vacuum state $|{\phi_0}\rangle.$
We observe that
\begin{equation}
\partial_i\Pi_{i}^{\gamma}({\bf{r}}) = \sum_{{\bf{k}}}
b_{Q}^{\gamma}({\bf{k}})
\;e^{+i{\bf{k}}\cdot{\bf{r}}}\;,
\label{eq:dipi}
\end{equation}
so that $\partial_i\Pi_{i}^{\gamma}({\bf{r}})|{\phi_0}\rangle = 0,$ and
\begin{equation}
[b_Q^a({\bf{k}}),\,{\cal{A}}_1]|{\phi_0}\rangle = -J_{0}^{a}({\bf{k}})
|{\phi_0}\rangle\;.
\label{eq:dipifermi}
\end{equation}
We might expect that the simple choice ${\Psi}={\Psi}_{0}=
\exp({\cal{A}}_1)$ would solve
Eq.~(\ref{eq:psicom}), but  that expectation is not fulfilled.
This is due to the fact that the  commutator
$[b_Q^a({\bf{k}}),\,{\cal{A}}_1]$
does not commute with ${\cal{A}}_1$.  The expression
$\partial_i\Pi^{\gamma}_{i}({\bf{r}}),$ when it arises in the midst of
an extended sequence of operator-valued factors, does not act on the
Fermi vacuum state, and does not vanish.  On its way to the extreme right
of the expression, where it ultimately acts on $|{\phi_0}\rangle$ and
vanishes, $\partial_i\Pi^{\gamma}_{i}({\bf{r}})$ produces extra terms
as it commutes with ${\cal{A}}_1$'s; and that fact disqualifies
${\Psi}_{0}$ as a solution of Eq.~(\ref{eq:psicom}). \bigskip

To address the problems that arise because ${\cal{A}}_1$ and
$[b_Q^a({\bf{k}}),\,{\cal{A}}_1]$ do not commute,
we make the following modifications. First, we replace
${\exp({\cal{A}}_1)}$ with ${\|}\,\exp({\cal{A}}_1)\,{\|}$,
where the ${\|}{\cal{O}}{\|}$
designates a variety of `normal order' in which all functionals
of momenta, ${\cal{F}}[\Pi_i]$, appear to the right of all
functionals of gauge fields, ${\cal{F}}[A_i]$.
For example, in the $n^{th}$ term of
${\|}\,\exp({\cal{A}}_1)\,{\|}$
the product ${\|}\,{(\cal{A}}_1)^n\,{\|}$
represents
\begin{equation}
{\|}\,{(\cal{A}}_1)^n\,{\|}=(ig)^n
\int{\sf D}(1,\cdots,n)\,\psi^{{\alpha_1}}_{(1)i_1}(1)\;
\psi^{{\alpha_2}}_{(1)i_2}(2)\;\cdots
\psi^{{\alpha_n}}_{(1)i_n}(n)\;
\Pi_{i_1}^{\alpha_1}(1)\;
\Pi_{i_2}^{\alpha_2}(2)\;\cdots\;
\Pi_{i_n}^{\alpha_n}(n)\;,
\label{eq:overn}
\end{equation}
where ${\sf D}(1,\cdots,n)$ denotes
$d{\bf r}_{1}\cdots d{\bf r}_{n},$ and the integer argument
$n$ in the ${\psi}$'s and ${\Pi}$'s represents ${\bf r}_{n}.$
This `normal order' has the effect that,
in $[b_Q^a({\bf{k}}),\,{\|}\,\exp({\cal{A}}_1)\,{\|}\,],$
the $\partial_i\Pi_i^\gamma({\bf{r}})$
produced by an integration by parts, appears among the
${\Pi}$'s and can annihilate the $|{\phi_0}\rangle$ directly, moving
only through other ${\Pi}$'s (or functionals of ${\Pi}$'s)  with which
it commutes.  The $(-g)\,f^{abc}\,{\int}\,d{\bf r}\,e^{-i{\bf{k}}
\cdot{\bf{r}}}\,[a^{b}_{i}({\bf{r}})+x^{b}_{i}({\bf{r}})]$ needed as part of
$-J_{0}^{a}({\bf{k}})$
is created to the left of ${\|}\,\exp({\cal{A}}_1)\,{\|},$
as required; but the remaining $\Pi_{i}^{c}({\bf{r}})$
appears to the right of all the ${\psi}$'s, with which
it {\em does not commute.}  As $\Pi_{i}^{c}({\bf{r}})$ moves to the left,
to constitute the
complete $-J_{0}^{a}({\bf{k}})$,  it
generates unwanted contributions that disqualify even
${\|}\,\exp({\cal{A}}_1)\,{\|}$ as a solution of  Eq.~(\ref{eq:psicom}).
To compensate for the failure of ${\|}\,\exp({\cal{A}}_1)\,{\|}$
to satisfy Eq.~(\ref{eq:psicom}), we extend ${\cal{A}}_1$ so
that it is only the first term in the infinite operator-valued
series ${\cal{A}},$ given by ${\cal{A}}=\sum_{n=1}^{\infty}{\cal{A}}_n$,
where the ${\cal{A}}_n$ with $n>1$ will be given later in this section.
$\Psi$ is then given as
\begin{equation}
{\Psi}={\|}\,\exp({\cal{A}})\,{\|}\;.
\label{eq:Apsi2}
\end{equation}
All the ${\cal A}_{n}$ will consist of
functionals of the gauge field, $A_{i}^{\alpha}({\bf{r}})$, multiplied by
a {\em single} momentum, $\Pi_{i}^{\gamma}({\bf{r}}),$ so that it
becomes useful to express all the ${\cal A}_{n}$ as
\begin{equation}
{\cal{A}}_{n}=ig^n{\int}d{\bf{r}}\;
{\cal{A}}_{(n)i}^{\gamma}({\bf{r}})\;
\Pi_i^{\gamma}({\bf{r}})\;.
\label{eq:Adens}
\end{equation}
Eq.~(\ref{eq:Adens}) is useful as a definition of
${\cal{A}}_{(n)i}^{\gamma}({\bf{r}}),$
and as an identification of this quantity as a functional
of gauge fields only --- canonical momenta are never
included in ${\cal{A}}_{(n)i}^{\gamma}({\bf{r}})$.
\bigskip

In Eq.~(\ref{eq:psicom}) we can now replace
$\Psi$ with Eq.~(\ref{eq:Apsi2}) to give
\begin{equation}
[b_{Q}^{a}({\bf{k}}),\,{\|}\,\exp({\cal A})\,
{\|}\,]\,+\,J_{0}^{a}({\bf{k}})
{\|}\,\exp({\cal A})\,{\|}\approx 0\;,
\label{eq:psicom1b}
\end{equation}
where the symbol $\approx$ is used to indicate that we have suppressed
the state $|\phi_0\rangle,$ that should appear on the right of all
operator products. We can expand Eq.~(\ref{eq:psicom1b})
into the form
\begin{eqnarray}
{\|}\,[b_{Q}^{a}({\bf{k}}),\,
\exp(\sum_{n=2}^\infty{\cal A}_n)\,]\,
\exp({\cal A}_1)\,{\|}\,
&+&\,{\|}\,[b_{Q}^{a}({\bf{k}}),\,\exp({\cal A}_1)\,]\,
\exp(\sum_{n=2}^\infty{\cal A}_n)\,{\|}
\nonumber\\
&+&J_{0}^{a}({\bf{k}})
{\|}\,\exp({\cal A})\,{\|}\approx 0\;,
\label{eq:psicom1c}
\end{eqnarray}
where the ${\|}\;\;\;{\|}$-ordering
eliminates further contributions from the Baker-Hausdorff-Campbell formula.
Since $[b_{Q}^{a}({\bf{k}}),\,{\|}\,\exp({\cal A})\,{\|}\,]
={\|}\,[b_{Q}^{a}({\bf{k}}),\,{\cal A}\,]\,
\exp({\cal A})\,{\|}$, we can use Eq.~(\ref{eq:thetabqcom}) to
eliminate $J_{0}^{a}({\bf{k}})
{\|}\,\exp({\cal A})\,{\|}$ and to rewrite Eq.~(\ref{eq:psicom1c}) as
\begin{equation}
{\|}\,[b_{Q}^{a}({\bf{k}}),\,
\sum_{n=2}^\infty{\cal A}_n\,]\exp({\cal A})\,{\|}\,
-\,{\|}\,g\,f^{a\beta\gamma}\int d{\bf{r}}\,e^{-i{\bf{k\cdot r}}}
[a^{\beta}_{i}({\bf{r}})+x^{\beta}_{i}({\bf{r}})]\,
[\exp({\cal A}),\,\Pi_{i}^{\gamma}({\bf r})]\,{\|}\approx 0\;.
\label{eq:psicom1f}
\end{equation}
We can also use
$[{\|}\,\exp({\cal A})\,{\|},\,\Pi_{i}^{\gamma}({\bf r})]
={\|}\,[{\cal A},\,\Pi_{i}^{\gamma}({\bf r})]\,
\exp({\cal A})\,{\|}$, to give\footnote{in an
expression ${\|}\,[\omega,\xi\,]\,\zeta\,{\|},$
the commutator is always to be evaluated {\em before} the double-bar
ordering is imposed.}
\begin{equation}
{\|}\,\{\,[b_{Q}^{a}({\bf{k}}),\,
\sum_{n=2}^\infty{\cal A}_n\,]\,
-\,g\,f^{a\beta\gamma}\int d{\bf{r}}\,e^{-i{\bf{k\cdot r}}}
[a^{\beta}_{i}({\bf{r}})+x^{\beta}_{i}({\bf{r}})]\,
[{\cal A},\,\Pi_{i}^{\gamma}({\bf r}_1)]\,\}\,\exp({\cal A})
\,{\|}\approx 0\;.
\label{eq:psicom1h}
\end{equation}
After simplifying  Eq.~(\ref{eq:psicom1h}) and expanding, we establish
\begin{equation}
[b_{Q}^{a}({\bf{k}}),\,\sum_{n=2}^\infty{\cal A}_n\,]\,
-\,g\,f^{a\beta\gamma}\int d{\bf{r}}\,e^{-i{\bf{k\cdot r}}}
\,[a^{\beta}_{i}({\bf{r}})+x^{\beta}_{i}({\bf{r}})]\,
[\sum_{n=1}^\infty{\cal A}_n,\,\Pi_{i}^{\gamma}({\bf r})]\approx 0\;,
\label{eq:psicom1j}
\end{equation}
as a sufficient condition for the validity of Eq.~(\ref{eq:psicom1h}).
We now rewrite the limit of the sum in the second term,
$\sum_{n=1}^\infty{\cal A}_n\Rightarrow\sum_{n=2}^\infty{\cal A}_{n-1}$,
to give
\begin{equation}
[b_{Q}^{a}({\bf{k}}),\,
\sum_{n=2}^\infty{\cal A}_n\,]\,
-\,g\,f^{a\beta\gamma}\int d{\bf{r}}\,e^{-i{\bf{k\cdot r}}}
\,[a^{\beta}_{i}({\bf{r}}_1)+x^{\beta}_{i}({\bf{r}})]\,
[\sum_{n=2}^\infty{\cal A}_{n-1},\,\Pi_{i}^{\gamma}({\bf r})]\approx 0\;.
\label{eq:psicom1k}
\end{equation}
Requiring Eq.~(\ref{eq:psicom1k}) to hold for all values of $g,$
we obtain the  the recursion relation
\begin{equation}
[b_Q^a({\bf{k}}),\,{\cal{A}}_n]\approx g\,f^{a\beta\gamma}
{\int}d{\bf r}\,e^{-i{\bf{k\cdot r}}}\,
[a^\beta_i({\bf{r}})+x^\beta_i({\bf{r}})]\,
[{\cal{A}}_{n-1},\,
\Pi^\gamma_i({\bf{r}})]\;,
\label{eq:recrel}
\end{equation}
which holds for $n>1$, because  ${\cal{A}}_0=0$. \bigskip

We have been able to construct the first six terms of the
${\cal{A}}$ series and to confirm their consistency with
Eq.~(\ref{eq:recrel}). Because of the structural regularity of
${\cal{A}}_1$ - ${\cal{A}}_6$, we can also infer the form of the general
${\cal{A}}_n.$  The expressions for these ${\cal{A}}_n$ are most
easily given in a partially recursive way, in terms
of the ${\cal{A}}_{(n)i}^{\gamma}({\bf{r}})$ previously defined in
Eq.~(\ref{eq:Adens}). The definition of each ${\cal{A}}_n$ (with $n>1$)
contains references to ${\cal{A}}_{(n^\prime)i}^{\gamma}({\bf{r}})$ with
$n^\prime <n,$ and, in turn, together with Eq.~(\ref{eq:Adens}),
defines the new ${\cal{A}}_{(n)i}^{\gamma}({\bf{r}}).$ The terms in the
${\cal{A}}$ series are given by Eq.~(\ref{eq:defscra1}) and by
\begin{eqnarray}
&&{\cal{A}}_2={\textstyle\frac{ig^2}{2}}{\int}d{\bf r}\;
\psi^{\gamma}_{(2)i}({\bf{r}})\;\Pi^\gamma_i({\bf{r}})
\nonumber\\
&+&ig^2\,f^{\alpha\beta\gamma}
{\int}d{\bf r}\;
{\textstyle\frac{\partial_j}{\partial^2}}
[{\cal{A}}_{(1)j}^\alpha({\bf{r}})]\;
a^{\beta}_{i}({\bf{r}})\;
\Pi^\gamma_i({\bf{r}})\;,
\label{eq:Atwolckh}
\end{eqnarray}
\begin{eqnarray}
&&{\cal{A}}_3={\textstyle\frac{ig^3}{3!}}
{\int}d{\bf r}\;\psi^{\gamma}_{(3)i}({\bf{r}})\;
\Pi^\gamma_i({\bf{r}})
\nonumber\\
&+&ig^3\,f^{\alpha\beta\gamma}
{\int}d{\bf r}\;
{\textstyle\frac{\partial_j}{\partial^2}}
[{\cal{A}}_{(2)j}^\alpha({\bf{r}})]
\;a^{\beta}_{i}({\bf{r}})\;\Pi^\gamma_i({\bf{r}})
\nonumber\\
&+&ig^3\,f^{\alpha\beta\gamma}
{\int}d{\bf r}\;
{\textstyle\frac{\partial_j}{\partial^2}}[
{\cal{A}}_{(1)j}^\alpha({\bf{r}})]\;
[\delta_{ik}-{\textstyle\frac{1}{2}}
{\textstyle\frac{\partial_i\partial_k}{\partial^2}}]
{\cal{A}}_{(1)k}^\beta({\bf{r}})\;
\Pi_{i}^{\gamma}({\bf{r}})\;,
\label{eq:Athreelckh}
\end{eqnarray}
\begin{eqnarray}
{\cal{A}}_4&=&{\textstyle\frac{ig^4}{4!}}
{\int}d{\bf r}\;\psi^{\gamma}_{(4)i}({\bf{r}})\;
\Pi^\gamma_i({\bf{r}})
\nonumber\\
&+&ig^4\,f^{\alpha\beta\gamma}
{\int}d{\bf r}\,
{\textstyle\frac{\partial_j}{\partial^2}}[
{\cal{A}}_{(3)j}^{\alpha}({\bf{r}})]\,
a_i^\beta({\bf{r}})\Pi^\gamma_i({\bf{r}})
\nonumber\\
&+&ig^4\,f^{\alpha\beta\gamma}{\int}d{\bf r}\;
{\textstyle\frac{\partial_j}{\partial^2}}
[{\cal{A}}_{(2)j}^{\alpha}({\bf{r}})]\,
[\delta_{ik}-{\textstyle\frac{1}{2}}
{\textstyle\frac{\partial_i\partial_k}{\partial^2}}]\;
{\cal{A}}_{(1)k}^{\beta}({\bf{r}})\,\Pi_i^\gamma({\bf{r}})
\nonumber\\
&+&ig^4\,f^{\alpha\beta\gamma}{\int}d{\bf r}\;
{\textstyle\frac{\partial_j}{\partial^2}}
[{\cal{A}}_{(1)j}^{\alpha}({\bf{r}})]\,
[\delta_{ik}-{\textstyle\frac{1}{2}}
{\textstyle\frac{\partial_i\partial_k}{\partial^2}}]\;
{\cal{A}}_{(2)k}^{\beta}({\bf{r}})\,\Pi_i^\gamma({\bf{r}})
\nonumber\\
&+&{\textstyle\frac{ig^4}{2}}\,f^{\alpha\beta b}
\,f^{b\delta\gamma}{\int}d{\bf r}\,
{\textstyle\frac{\partial_j}{\partial^2}}
[{\cal{A}}_{(1)j}^{\alpha}({\bf{r}})]
{\textstyle\frac{\partial_k}{\partial^2}}
[{\cal{A}}_{(1)k}^{\delta}({\bf{r}})]
\;a^{\beta}_{i}({\bf{r}})
\,\Pi^\gamma_i({\bf{r}})\;,
\label{eq:Afourlckh}
\end{eqnarray}
\begin{eqnarray}
{\cal{A}}_5&=&{\textstyle\frac{ig^5}{5!}}
{\int}d{\bf r}\;\psi^{\gamma}_{(5)i}({\bf{r}})\;
\Pi^\gamma_i({\bf{r}})
\nonumber\\
&+&ig^5\,f^{\alpha\beta\gamma}
{\int}d{\bf r}\,
{\textstyle\frac{\partial_j}{\partial^2}}[
{\cal{A}}_{(4)j}^{\alpha}({\bf{r}})]\,
a_i^\beta({\bf{r}})\Pi^\gamma_i({\bf{r}})
\nonumber\\
&+&ig^5\,f^{\alpha\beta\gamma}{\int}d{\bf r}\;
{\textstyle\frac{\partial_j}{\partial^2}}
[{\cal{A}}_{(3)j}^{\alpha}({\bf{r}})]\,
[\delta_{ik}-{\textstyle\frac{1}{2}}
{\textstyle\frac{\partial_i\partial_k}{\partial^2}}]\;
{\cal{A}}_{(1)k}^{\beta}({\bf{r}})\,\Pi_i^\gamma({\bf{r}})
\nonumber\\
&+&ig^5\,f^{\alpha\beta\gamma}{\int}d{\bf r}\;
{\textstyle\frac{\partial_j}{\partial^2}}
[{\cal{A}}_{(2)j}^{\alpha}({\bf{r}})]\,
[\delta_{ik}-{\textstyle\frac{1}{2}}
{\textstyle\frac{\partial_i\partial_k}{\partial^2}}]\;
{\cal{A}}_{(2)k}^{\beta}({\bf{r}})\,\Pi_i^\gamma({\bf{r}})
\nonumber\\
&+&ig^5\,f^{\alpha\beta\gamma}{\int}d{\bf r}\;
{\textstyle\frac{\partial_j}{\partial^2}}
[{\cal{A}}_{(1)j}^{\alpha}({\bf{r}})]\,
[\delta_{ik}-{\textstyle\frac{1}{2}}
{\textstyle\frac{\partial_i\partial_k}{\partial^2}}]\;
{\cal{A}}_{(3)k}^{\beta}({\bf{r}})\,\Pi_i^\gamma({\bf{r}})
\nonumber\\
&+&{\textstyle\frac{ig^5}{2}}\,f^{\alpha\beta b}
\,f^{b\delta\gamma}{\int}d{\bf r}\,
{\textstyle\frac{\partial_j}{\partial^2}}
[{\cal{A}}_{(2)j}^{\alpha}({\bf{r}})]
{\textstyle\frac{\partial_k}{\partial^2}}
[{\cal{A}}_{(1)k}^{\delta}({\bf{r}})]
\;a^{\beta}_{i}({\bf{r}})
\,\Pi^\gamma_i({\bf{r}})
\nonumber\\
&+&{\textstyle\frac{ig^5}{2}}\,f^{\alpha\beta b}
\,f^{b\delta\gamma}{\int}d{\bf r}\,
{\textstyle\frac{\partial_j}{\partial^2}}
[{\cal{A}}_{(1)j}^{\alpha}({\bf{r}})]
{\textstyle\frac{\partial_k}{\partial^2}}
[{\cal{A}}_{(2)k}^{\delta}({\bf{r}})]
\;a^{\beta}_{i}({\bf{r}})
\,\Pi^\gamma_i({\bf{r}})
\nonumber\\
&+&{\textstyle\frac{ig^5}{2}}\,f^{\alpha\beta b}
\,f^{b\delta\gamma}{\int}d{\bf r}\;
{\textstyle\frac{\partial_j}{\partial^2}}
[{\cal{A}}_{(1)j}^\alpha({\bf{r}})]\;
{\textstyle\frac{\partial_k}{\partial^2}}
[{\cal{A}}_{(1)k}^\delta({\bf{r}})]\;
[\delta_{il}-{\textstyle\frac{2}{3}}
{\textstyle\frac{\partial_i\partial_l}{\partial^2}}]
{\cal{A}}_{(1)l}^\beta({\bf{r}})\;
\Pi_{i}^{\gamma}({\bf{r}})\;,
\label{eq:Afivelckh}
\end{eqnarray}
and
\begin{eqnarray}
{\cal{A}}_6&=&{\textstyle\frac{ig^6}{6!}}
{\int}d{\bf r}\;\psi^{\gamma}_{(6)i}({\bf{r}})\;
\Pi^\gamma_i({\bf{r}})
\nonumber\\
&+&ig^6\,f^{\alpha\beta\gamma}
{\int}d{\bf r}\,
{\textstyle\frac{\partial_j}{\partial^2}}[
{\cal{A}}_{(5)j}^{\alpha}({\bf{r}})]\,
a_i^\beta({\bf{r}})\Pi^\gamma_i({\bf{r}})
\nonumber\\
&+&ig^6\,f^{\alpha\beta\gamma}{\int}d{\bf r}\;
{\textstyle\frac{\partial_j}{\partial^2}}
[{\cal{A}}_{(4)j}^{\alpha}({\bf{r}})]\,
[\delta_{ik}-{\textstyle\frac{1}{2}}
{\textstyle\frac{\partial_i\partial_k}{\partial^2}}]\;
{\cal{A}}_{(1)k}^{\beta}({\bf{r}})\,\Pi_i^\gamma({\bf{r}})
\nonumber\\
&+&ig^6\,f^{\alpha\beta\gamma}{\int}d{\bf r}\;
{\textstyle\frac{\partial_j}{\partial^2}}
[{\cal{A}}_{(3)j}^{\alpha}({\bf{r}})]\,
[\delta_{ik}-{\textstyle\frac{1}{2}}
{\textstyle\frac{\partial_i\partial_k}{\partial^2}}]\;
{\cal{A}}_{(2)k}^{\beta}({\bf{r}})\,\Pi_i^\gamma({\bf{r}})
\nonumber\\
&+&ig^6\,f^{\alpha\beta\gamma}{\int}d{\bf r}\;
{\textstyle\frac{\partial_j}{\partial^2}}
[{\cal{A}}_{(2)j}^{\alpha}({\bf{r}})]\,
[\delta_{ik}-{\textstyle\frac{1}{2}}
{\textstyle\frac{\partial_i\partial_k}{\partial^2}}]\;
{\cal{A}}_{(3)k}^{\beta}({\bf{r}})\,\Pi_i^\gamma({\bf{r}})
\nonumber\\
&+&ig^6\,f^{\alpha\beta\gamma}{\int}d{\bf r}\;
{\textstyle\frac{\partial_j}{\partial^2}}
[{\cal{A}}_{(1)j}^{\alpha}({\bf{r}})]\,
[\delta_{ik}-{\textstyle\frac{1}{2}}
{\textstyle\frac{\partial_i\partial_k}{\partial^2}}]\;
{\cal{A}}_{(4)k}^{\beta}({\bf{r}})\,\Pi_i^\gamma({\bf{r}})
\nonumber\\
&+&{\textstyle\frac{ig^6}{2}}\,f^{\alpha\beta b}
\,f^{b\delta\gamma}{\int}d{\bf r}\,
{\textstyle\frac{\partial_j}{\partial^2}}
[{\cal{A}}_{(3)j}^{\alpha}({\bf{r}})]
{\textstyle\frac{\partial_k}{\partial^2}}
[{\cal{A}}_{(1)k}^{\delta}({\bf{r}})]
\;a^{\beta}_{i}({\bf{r}})
\,\Pi^\gamma_i({\bf{r}})
\nonumber\\
&+&{\textstyle\frac{ig^6}{2}}\,f^{\alpha\beta b}
\,f^{b\delta\gamma}{\int}d{\bf r}\,
{\textstyle\frac{\partial_j}{\partial^2}}
[{\cal{A}}_{(2)j}^{\alpha}({\bf{r}})]
{\textstyle\frac{\partial_k}{\partial^2}}
[{\cal{A}}_{(2)k}^{\delta}({\bf{r}})]
\;a^{\beta}_{i}({\bf{r}})
\,\Pi^\gamma_i({\bf{r}})
\nonumber\\
&+&{\textstyle\frac{ig^6}{2}}\,f^{\alpha\beta b}
\,f^{b\delta\gamma}{\int}d{\bf r}\,
{\textstyle\frac{\partial_j}{\partial^2}}
[{\cal{A}}_{(1)j}^{\alpha}({\bf{r}})]
{\textstyle\frac{\partial_k}{\partial^2}}
[{\cal{A}}_{(3)k}^{\delta}({\bf{r}})]
\;a^{\beta}_{i}({\bf{r}})
\,\Pi^\gamma_i({\bf{r}})
\nonumber\\
&+&{\textstyle\frac{ig^6}{2}}\,f^{\alpha\beta b}
\,f^{b\delta\gamma}{\int}d{\bf r}\;
{\textstyle\frac{\partial_j}{\partial^2}}
[{\cal{A}}_{(2)j}^\alpha({\bf{r}})]\;
{\textstyle\frac{\partial_k}{\partial^2}}
[{\cal{A}}_{(1)k}^\delta({\bf{r}})]\;
[\delta_{il}-{\textstyle\frac{2}{3}}
{\textstyle\frac{\partial_i\partial_l}{\partial^2}}]
{\cal{A}}_{(1)l}^\beta({\bf{r}})\;
\Pi_{i}^{\gamma}({\bf{r}})
\nonumber\\
&+&{\textstyle\frac{ig^6}{2}}\,f^{\alpha\beta b}
\,f^{b\delta\gamma}{\int}d{\bf r}\;
{\textstyle\frac{\partial_j}{\partial^2}}
[{\cal{A}}_{(1)j}^\alpha({\bf{r}})]\;
{\textstyle\frac{\partial_k}{\partial^2}}
[{\cal{A}}_{(2)k}^\delta({\bf{r}})]\;
[\delta_{il}-{\textstyle\frac{2}{3}}
{\textstyle\frac{\partial_i\partial_l}{\partial^2}}]
{\cal{A}}_{(1)l}^\beta({\bf{r}})\;
\Pi_{i}^{\gamma}({\bf{r}})
\nonumber\\
&+&{\textstyle\frac{ig^6}{2}}\,f^{\alpha\beta b}
\,f^{b\delta\gamma}{\int}d{\bf r}\;
{\textstyle\frac{\partial_j}{\partial^2}}
[{\cal{A}}_{(1)j}^\alpha({\bf{r}})]\;
{\textstyle\frac{\partial_k}{\partial^2}}
[{\cal{A}}_{(1)k}^\delta({\bf{r}})]\;
[\delta_{il}-{\textstyle\frac{2}{3}}
{\textstyle\frac{\partial_i\partial_l}{\partial^2}}]
{\cal{A}}_{(2)l}^\beta({\bf{r}})\;
\Pi_{i}^{\gamma}({\bf{r}})
\nonumber\\
&+&{\textstyle\frac{ig^6}{3!}}\,f^{\alpha\beta b}
\,f^{b\mu c}\,f^{c\delta\gamma}{\int}d{\bf r}\;
{\textstyle\frac{\partial_j}{\partial^2}}
[{\cal{A}}_{(1)j}^\alpha({\bf{r}})]\;
{\textstyle\frac{\partial_k}{\partial^2}}
[{\cal{A}}_{(1)k}^\mu({\bf{r}})]\;
{\textstyle\frac{\partial_l}{\partial^2}}
[{\cal{A}}_{(1)l}^\delta({\bf{r}})]\;
a^{\beta}_{i}({\bf{r}})\;
\Pi_{i}^{\gamma}({\bf{r}})\;.
\label{eq:Asixlckh}
\end{eqnarray}

To arrive at a form for ${\cal{A}}_n$ for arbitrary $n,$ it is convenient to
define $\overline{{\cal{A}}_{i}^{\gamma}}({\bf{r}})$ by Eq.~(\ref{eq:Adens})
and by
\begin{equation}
\overline{{\cal{A}}_{i}^{\gamma}}({\bf{r}})=
\sum_{n=1}^\infty g^n
{\cal{A}}_{(n)i}^{\gamma}({\bf{r}})\;,
\end{equation}
so that
\begin{equation}
{\cal{A}}=
i{\int}d{\bf{r}}\;
\overline{{\cal{A}}_{i}^{\gamma}}({\bf{r}})\;
\Pi_i^{\gamma}({\bf{r}})\;.
\label{eq:Awhole}
\end{equation}
We also define
\begin{equation}
\overline{{\cal{B}}_{(\eta) i}^{\beta}}({\bf{r}})=
\{a_i^\beta({\bf{r}})+\,[\delta_{ij}-{\textstyle\frac{\eta}{\eta+1}}
{\textstyle\frac{\partial_i\partial_j}{\partial^2}}]
\overline{{\cal{A}}_{j}^{\beta}}({\bf{r}})\,\}\;,
\label{eq:calB1b}
\end{equation}
and
\begin{equation}
{\cal{M}}_{(\eta)}^{\vec{\alpha}}({\bf{r}})
\equiv\prod_{m=1}^\eta
{\textstyle\frac{\partial_{j}}{\partial^2}}
[\overline{{\cal{A}}_{j}^{\alpha[m]}}({\bf{r}})]
={\textstyle\frac{\partial_j}{\partial^2}}
[\overline{{\cal{A}}_{j}^{\alpha[1]}}({\bf{r}})]
{\textstyle\frac{\partial_l}{\partial^2}}
[\overline{{\cal{A}}_{l}^{\alpha[2]}}({\bf{r}})]
\cdots{\textstyle\frac{\partial_w}{\partial^2}}
[\overline{{\cal{A}}_{w}^{\alpha[\eta-1]}}({\bf{r}})]
{\textstyle\frac{\partial_k}{\partial^2}}
[\overline{{\cal{A}}_{k}^{\alpha[\eta]}}({\bf{r}})]\,,
\end{equation}
where $\eta$ represents the same integer-valued parameter originally introduced
in
Eq.~(\ref{eq:bookaiQ}),  that we will now observe to be necessary for matching
each $\overline{{\cal{B}}_{(\eta) i}^{\beta}}({\bf{r}})$ with its
corresponding ${\cal{M}}_{(\eta)}^{\vec{\alpha}}({\bf{r}})$ in the integral
equation for $\overline{{\cal{A}}_{i}^{\gamma}}({\bf{r}})$ given below.
We now formulate this non-linear integral equation for
$\overline{{\cal{A}}_{i}^{\alpha}}({\bf{r}})$ as
\begin{eqnarray}
&&{\cal{A}}=\sum_{\eta=1}^\infty
{\textstyle\frac{ig^\eta}{\eta!}}{\int}d{\bf r}\;
\psi^{\gamma}_{(\eta)i}({\bf{r}})\;\Pi^\gamma_i({\bf{r}})
\nonumber\\
&&+ig\;f^{\vec{\alpha}\beta\gamma}_{(1)}\,
{\int}d{\bf r}
{\cal{M}}_{(1)}^{\vec{\alpha}}({\bf{r}})\;
\overline{{\cal{B}}_{(1) i}^{\beta}}({\bf{r}})\;
\Pi^\gamma_i({\bf{r}})
\nonumber\\
&&+{\textstyle\frac{ig^2}{2}}
\;f^{\vec{\alpha}\beta\gamma}_{(2)}\,
{\int}d{\bf r}\,
{\cal{M}}_{(2)}^{\vec{\alpha}}({\bf{r}})\;
\overline{{\cal{B}}_{(2) i}^{\beta}}({\bf{r}})\;
\Pi^\gamma_i({\bf{r}})
\nonumber\\
&&+{\textstyle\frac{ig^3}{3!}}
\;f^{\vec{\alpha}\beta\gamma}_{(3)}\,
{\int}d{\bf r}\,{\cal{M}}_{(3)}^{\vec{\alpha}}({\bf{r}})\,
\overline{{\cal{B}}_{(3) i}^{\beta}}({\bf{r}})\,
\Pi^\gamma_i({\bf{r}})
\nonumber\\
&&+\cdots
\nonumber\\
&&+{\textstyle\frac{ig^\eta}{\eta!}}
\,f^{\vec{\alpha}\beta\gamma}_{(\eta)}\,
{\int}d{\bf r}\,{\cal{M}}_{(\eta)}^{\vec{\alpha}}({\bf{r}})\,
\overline{{\cal{B}}_{(\eta) i}^{\beta}}({\bf{r}})\,
\Pi^\gamma_i({\bf{r}})
\nonumber\\
&&+\cdots\;\;\;\;,
\label{eq:inteq1}
\end{eqnarray}
or, more succinctly,
\begin{equation}
{\cal{A}}=\sum_{\eta=1}^\infty
{\textstyle\frac{ig^\eta}{\eta!}}{\int}d{\bf r}\;
\{\,\psi^{\gamma}_{(\eta)i}({\bf{r}})\,+
\,f^{\vec{\alpha}\beta\gamma}_{(\eta)}\,
{\int}d{\bf r}\,{\cal{M}}_{(\eta)}^{\vec{\alpha}}({\bf{r}})\,
\overline{{\cal{B}}_{(\eta) i}^{\beta}}({\bf{r}})\,\}\;
\Pi^\gamma_i({\bf{r}})\;.
\label{eq:inteq2}
\end{equation}
We observe that the leading terms
of the perturbative solution of
Eq.~(\ref{eq:inteq1}) --- or equivalently
Eq.~(\ref{eq:inteq2}) --- agree with ${\cal{A}}_1$ - ${\cal{A}}_6,$
the explicit forms given in
Eqs.~(\ref{eq:defscra1}) and (\ref{eq:Atwolckh})-(\ref{eq:Asixlckh}).
We have confirmed that the
entire perturbative series --- ${\cal{A}}_{n}$ for arbitrary
$n$ ---  correctly satisfy the recursion relation given in
Eq.~(\ref{eq:recrel});
but we have not yet established that fact with
complete rigor in so far as concerns its extension beyond $n=6$. Previously
in this paper, we have shown Eq.~(\ref{eq:recrel}) to be a sufficient
condition for the implementation of Gauss's law.\bigskip

$|\nu_0\rangle$ is not the only state that implements Gauss's law.  Any state
$|\nu_{\bf{k}}\rangle ={\Psi}\,{\Xi}\,a_{s}^{a\dagger}({\bf{k}})|0\rangle$ or
$|\nu_{{\bf{k}}_1\cdots{\bf{k}}_i}\rangle ={\Psi}\,{\Xi}
\,a_{s_1}^{a_1\dagger}({\bf{k}}_1)\cdots\,a_{s_i}^{a_i\dagger}({\bf{k}}_i)
|0\rangle$,  where $a_{s}^{a\dagger}({\bf{k}})$ creates a transversely
polarized gluon,  is annihilated by the Gauss's law operator ${\cal G}.$
The important question then arises: What changes occur in the S-matrix when
the states $|\nu_{{\bf{k}}_1\cdots{\bf{k}}_i}\rangle$
are substituted for the perturbative states
$|n_{{\bf{k}}_1\cdots{\bf{k}}_i}\rangle$ as incident
and scattered states? We will not discuss this question in
detail in this paper. But we observe
that, while it has been shown that there is no change in S-matrix elements
when the $|\nu\rangle$ and the corresponding
$|n\rangle$ states that they replace
are unitarily equivalent~\cite{khqedtemp,khelqed},
in {\em this} case, in which the transformation ${\Psi}$ is {\em not}
unitary, there {\em will} be changes in the S-matrix elements when the
$|\nu\rangle$ states are substituted for the $|n\rangle$ states.
These changes, however, do not appear in the lowest order of perturbation
theory, since the non-unitarity does not arise in the leading term of
${\Psi}.$  The most interesting possibility, of course is that the
contribution of ${\Psi}$ as a whole can be evaluated, and
the non-perturbative effect of Gauss's law on the S-matrix assessed. \bigskip

Another question to be addressed deals
with the non-perturbative solutions of
the non-linear Eq.~(\ref{eq:inteq1}) or equivalently
Eq.~(\ref{eq:inteq2}).  Further work is required
to clarify how these non-perturbative solutions are related to
the gauge sectors connected by the
large gauge transformations~\cite{raja}.\bigskip

This research was supported by the Department of Energy
under Grant No. DE-FG02-92ER40716.00.

\end{document}